\documentclass[preprint,12pt]{elsarticle}

\usepackage{amssymb}
\usepackage{todonotes}
\usepackage{natbib}
\usepackage{amsmath}
\usepackage{footnote}
\usepackage{hyperref}
\usepackage{float}

\journal{Computer Speech and Language}

\begin{document}

\begin{frontmatter}



\title{Combining a Context Aware Neural Network with a Denoising Autoencoder for Measuring String Similarities}


\author{Mehdi Ben Lazreg}
\author{Morten Goodwin}
\address{University of Agder, Department of Information and Communication Technologies, Grimstad, Norway}

\begin{abstract}
Measuring similarities between strings is central for many established and fast growing research areas including information retrieval, biology, and natural language processing. The traditional approach for string similarity measurements is to define a metric over a word space that quantifies and sums up the differences between characters in two strings. The state-of-the-art in the area has, surprisingly, not evolved much during the last few decades. The majority of the metrics are based on a simple comparison between character and character distributions without consideration for the context of the words. This paper proposes a string metric that encompasses similarities between strings based on (1) the character similarities between the words including. Non-Standard and standard spellings of the same words, and (2) the context of the words. Our proposal is a neural network composed of a denoising autoencoder and what we call a context encoder specifically designed to find similarities between the words based on their context. The experimental results show that the resulting metrics succeeds in 85.4\% of the cases in finding the correct version of a non-standard spelling among the closest words, compared to 63.2\% with the established Normalised-Levenshtein distance. Besides, we show that words used in similar context are with our approach calculated to be similar than words with different contexts, which is a desirable property missing in established string metrics. 
\end{abstract}

\begin{keyword}
String Metric \sep Neural network \sep Autoencoder \sep Context encoder



\end{keyword}

\end{frontmatter}


\section{Introduction}
\label{sec_introduction}

Measuring similarities between strings is an essential component for a large number of language and string processing tasks including information retrieval, natural biology, and natural language processing. A string metric is a function that quantifies the distance between two strings. The most widely known string metric is the edit distance, also known as the Levenshtein distance which represents the number of substitutions insertions, or deletions operation needed to transform one string to another \cite{levenshtein1966binary}. The fewer operations needed to go from one string to another, the more similar two strings are.

String metrics are key to the approximate string matching problem present in many fields. As an example, natural language processing needs automatic correction of spelling, and in bioinformatics similarities between DNA sequences is a crucial task. Both tasks are string approximation problems. Common to all string similarity metrics is that they are used to find matching patterns for a string that underwent some distortion including, but not limited to, misspellings, abbreviations, slang or irregularities in DNA sequences. The focus in this paper is a novel distance metric with some advantageous properties not present in other string metrics including considerations misspellings and non-standard usage of words, and including the string context. 

The evolution and distortion of languages is nothing new. However, a consequence of the global social media era is the non-standardization of languages, which means that the same phrase, and even the same word, can be communicated in a variety of ways within the same language. This evolution is a challenge for natural language processing as any data handling, such as classification of translation, becomes less formal. All natural language processing would be much easier if everyone wrote the same way --- which is unrealistic. A mitigation is to normalize non-standard words to a more standard format that is easier to handle.

o normalize misspellings, non-standard words and phrases, abbreviations, dialects, sociolects and other text variations (referred to here as non-standard spelling and non-standard words) three approaches other than string metrics are available in the literature. The first method is to view normalization of any non-standard spelling as a translation problem \cite{aw2006phrase}. One example is based on statistical tools that map the non-standard words with their English counterparts based on a probability distribution\cite{kobus2008normalizing}. Certainly, the translation method is a promising strategy, however, using a method designed to capture the complex relationship between two different languages in the context of word normalization is an ``overreach" given the strong relationship between the English words and their non-standard forms. A second approach for solving the non-standard spelling challenge is to consider it as plain spell checking. The method tries to correct the misspelled words based a probability model\cite{aw2006phrase}. The challenge with the latter is that distance between non-standard and standard spelling substantial. Hence, far from all normalizations can be viewed as corrections. Third, normalizing non-standard spelling can be viewed as a speech recognition problem\cite{aw2006phrase} \cite{pennell2011toward}. In this approach, the texts are regarded as a phonetic approximation of the correctly spelled message. What motivates this view is that many non-standard spellings are written based on their phonetic rather than their normative spelling. However, this view is also an oversimplification of the nature of the texts as it contains non-standard spellings that are not just phonetic spellings of the correct word. As examples, texts containing abbreviation (lol for laugh out loud), truncated words (n for and), and leetspeak (4ever for forever) not be handled by this approach.

This paper proposes a new method that maps each word in a vocabulary into a real vectors space. As a result, the distance between two words will be the distance between the vector representation of those words in the real vector space. The mapping between the word vocabulary and the real vector space must satisfy two premises. The first premise is that the distance between the vector representations of a non-standard word and its corrected form should be shorter than the distance between the non-standard word and any other unrelated known word. To achieve this constraint, the vector representation needs to spur positive correlations between a corrected spelling and \emph{every} possible non-standard spelling of that word, while minimizing correlation with all other words and \emph{their} non-standard spellings. The second premise is that the vector representation of a word should also be such as each word with similar meaning has similar representations. We assume that two words have a similar meaning when they are often used in the same context. The context of a word is the collection of words surrounding it. To obtain such a representation, we mixed a predictive word embedding methods with a denoising autoencoder\cite{bengio2009learning}. A denoising autoencoder is an artificial neural network that takes as input a data set, adds some noise to the data, then tries to reconstruct the initial data from the noisy version. By performing this reconstruction, the denoising autoencoder learns the feature present in the initial data in its hidden layer. In our approach, we consider the non-standard spellings to be the noisy versions of the corrected word forms. In a predictive word embedding method, Each word is represented by a real-valued vector learned based on the usage of words and their context. A neural network learns this real-valued vector representation in a way that minimizes the loss of predicting a word based on its context. This representation is in contrast to the representation in a bag of words model where, unless explicitly managed, different words have different representations, regardless of their use.

\section{Background}
\label{sec_background}

A string metric or string distance function defines a distance between every element of a set of strings $A$. Any distance function $d$ on $A$ must satisfy the following conditions for all $x, y,$ and $z \in A$ \cite{hazewinkel2013encyclopaedia}:
\begin{equation}
    d(x,y)\geq 0 \quad \text{non-negativity}
    \label{eq_non-negativity}
\end{equation} 
\begin{equation}
    d(x,y)=0 \iff x=y \quad \text{identity of indiscernibles}
    \label{eq_	identity}
\end{equation} 
\begin{equation}
    d(x,y)=d(y,x) \quad \text{symmetry}
    \label{eq_symmetry}
\end{equation} 
\begin{equation}
    d(x,z)\leq d(x,y)+d(y,z) \quad \text{triangle inequality}
    \label{eq_triangle}
\end{equation} 
Hence, the comparison between any two strings is larger or equal to 0 (Equation \ref{eq_non-negativity}), identical strings have distance 0 (Equation \ref{eq_	identity}), the distance between two strings is independent of whether the first is compared to the second or visa verse (Equation \ref{eq_symmetry}), and the distance between two strings is always equal to or smaller than including a third string in the measurements (Equation \ref{eq_triangle}). 

Over the years, several attempts at defining an all-encompassing string metric have been carried out. The most well known is the edit distance (Levenshtein distance) which has been one of the most widely used string comparison function since its introduction in 1965 \cite{levenshtein1966binary}. It counts the minimum number of operations (deletes, insert and substitute of a character)  required to transform one string to another.  It also assigns a cost to each operation. For example, if the weights assigns to the operation sis one, the distance between the words ``vector" and ``doctor" is two since only two substitutions are required for a transformation. The edit distance satisfies all the requirements as a distance function (equations \ref{eq_non-negativity},\ref{eq_    identity},\ref{eq_symmetry} and \ref{eq_triangle}). 

The edit distance is called a simple edit distance when all operations have the same cost and a general edit distance when operations have different costs. Other than that, the edit distance has four notable variants. First, a longest common subsequence (LCS) is when only insertions and deletions are allowed with cost one \cite{bakkelund2009lcs}. A second simplification is a variant that only allows substitution. In this case, the distance is called the Hamming distance \cite{hamming1950error}. Third, the Damerau-Levenshtein Distances adds the transposition of two adjacent characters to the operations allowed by the edit distance \cite{damerau1964technique}.  Finally, the episode distance allows only insertions that cost 1. The episode distance is not symmetric and does not satisfy Equation \ref{eq_symmetry}. Since insertions do not allow to transform a string $x$ to $y$, $d(x,y)$ is either $|y|-|x|$ or $\infty$.

In 1992, Ukkonen \cite{ukkonen1992approximate} introduced the q-gram distance. It is based on counting the number of occurrences of common q-grams (strings of length $q \in \mathbb{N}$) in each string, and the strings have a closer distance the more q-grams they have in common. The q-gram distance is not a metric function since it does not obey the identity of indiscernibles requirement (Equation \ref{eq_    identity}). 

Later, Kondrak \cite{kondrak2005n} developed the notion of N-gram distance in which he extended the edit and LCS distance to consider the deletions, insertions, and substitutions of N-grams. The use of N-grams enabled some new statical methods for string metrics originally from the field of samples and sets. The use of N-gram introduces the notion of statistical string metrics, which are metrics that measure the statistical properties of the compared strings. An example, the Sorensen-Dice coefficient was used as a metric to measure the similarity between two string \cite{sorensen1948method}\cite{dice1945measures}, initially a method used to compare the similarity between two samples. In the case of strings the coefficient is computed as follows:
\begin{equation}
    d(x,y)= \frac{2n_t}{n_x+n_y}
    \label{eq_sd}
\end{equation}
where $n_t$ is the number of character N-grams found in both strings, $n_x$ is the number of N-grams in string x and $n_y$ is the number of N-grams in string $y$.

The Jaccard index is another statistical method used to compare the similarity between two sample sets, including strings. It is calculated as one minus the quotient of shared N-grams by all observed N-grams in both strings.

Some vector similarity functions have been extended to include string similarity as well, of which the most notable is the string cosine similarity. It measures the cosine similarity between vector representations of two strings to be compared. For English words, the vectors have a size 26, one element for each character, and the number of occurrences of each character in each string.   

The use of machine learning techniques for vector representations of words has been around since 1986 thanks to the work of Rumelhart, Hinton, and Williams\cite{rumelhart1988learning}. The string similarity measurements are used as features in supervised natural language processing tasks to increase the performance of the classifier. More recently, a method called locally linear embedding was introduced. The method computes low-dimensional, neighborhood-preserving embedding of high dimensional input. The method is applied to generate a two-dimensional embedding of words that conserves their semantics\cite{roweis2000nonlinear}. Later, feedforward neural networks were used to generate a distributed vector representation of words\cite{bengio2003neural}. By predicting the next word giving the previous words in the context, the neural network learns a vector representation of the words in its hidden layer. The method is extended to take into consideration the surrounding words not only the previous words\cite{mikolov2013distributed}. In the same context, the feedforward neural network is replaced by a restricted Boltzmann machine to produce the vector representations\cite{mnih2007three}. A word vector representation variant learns for each word a low dimensional linear projection of the one-hot encoding of a word by incorporation the projection in the energy function of a restricted Boltzmann machine\cite{dahl2012training}\cite{hinton2009replicated}. Finally, GloVe is one of the most successful attempts at producing vector representations of words for string comparisons \cite{pennington2014glove}. GloVe learns a log-bi-linear model that combines the advantages of global matrix factorization and local context window to produce a vector representation of word based on the word count. A vector similarity measure such as the Euclidean distance, cosine similarity, or $L_1$ measure can then be used to measure the similarity between two strings.

\section{Word coding approach}
\label{sec_approch}
The objective of this research is to find a function $F$ that maps words into real vector space in such as a way the distance between two similar words (i.e., non-standard spellings of the same word or words used in the same context) will be the smallest distance between the corresponding mapping in the real vector space. To achieve this goal, $F$ needs to obey two constraints. The first constraint is that the distance in real vector space between the mapping of a word and its non-standard versions must be shorter than the distance between that word and non-standard versions of other words. The second constraint is that the distance in real vector space between the mapping of words with similar meanings must be shorter than the distance between words with dissimilar meanings. We define meaning by similar context: we assume that words used in the same context have a similar meaning. To mode the first constraint, we use a denoising autoencoder, and to model the second constraint, we introduce a context encoder.

The denoising autoencoder and the context encoder are explained in sections \ref{subsec_do} and \ref{subsec_co} respectively. The overall method is explained in section \ref{subsec_distance}. A summary of all the notation, parameters, and functions used in this section is summarized in Appendix \ref{apx_pandf}.

\subsection{Denoising autoencoder}
\label{subsec_do}
An autoencoder is an unsupervised learning algorithm based on artificial neural networks in which the target value is equal to the input\cite{bengio2009learning}. An autoencoder can in its simplest form be represented by a network composed of:
\begin{itemize}
  \item An input layer representing the feature vector of the input.
  \item A hidden layer that applies a non-linear transformation of the input.
  \item An output layer representing the target value or the label.
\end{itemize}

Suppose we have a training example $x$, the autoencoder tries to learn a function $id$ such that $id(x)\simeq x$, an approximation of the identity function. The identity function seems to be a trivial function to learn, however, if we put some constraints on the autoencoder, it can learn a function that captures features and structures in the data. For example, limiting the number of hidden units in the network to be less than the input units forces the network to learn a compressed representation of the input. Instead of copying the value of the input in the hidden layer, the network must learn which parts of the input is more important and leads to a better reconstruction. Adding noise to the input is another constraint that forces the autoencoder to learn the most features of the data. By reconstructing the data based on its noisy version, the autoencoder undoes the effect of the noise. Undoing the noisy effects can only be performed when the autoencoder learns the statistical dependencies between the inputs. For the latter example, the autoencoder is called a denoising autoencoder.

In our approach, we input the non-standard spelling to a denoising autoencoder and try to reconstruct the original word. Any non-standard spelling of a word can be seen as a noisy version of the original word. The aim is that the network should learn two essential features: (1) The relations between non-standard and standard word spellings, and (2) what separates the standard words. Both features should lie in the hidden layer, which is used to reconstruct the standard word from the non-standard spellings. 

The denoising autoencoder in our approach includes a vocabulary $A=\{a_1,a_2,...,a_r\}$ of $r$ words, which can be standard English\footnote{The approach is not limited to English. However, all our examples are from the English language.} words and non-standard variants of those words. $A$ consists of the following subsets: The standard words $C=\{c_1,c_2,...,c_m\} \subset A$, and the non-standard spelling for of every word $c_i \in C$ as $M_{c_i} \subset A$. 
We define an initialization function $v$ that transforms a word in $A$ into a vector of real numbers in $\mathbb{R}^n$. $v$ can be a function that performs a one-hot encoding of the words in $A$, or it can map each character in a word to a unique number, or assign a random vector to each word. $v$ can be presented by a $\mid A\mid \times n$ matrix of free parameters.


The input of the denoising autoencoder is the non-standard word spelling $m_j \in M_{c_i}$ corresponding to word $c_i$, and the output of the hidden layer is $h(v(m_j))$. The reconstructed word outputed from the autoencoder, $\boldsymbol{\tilde{c_i}}$ should ideally be equal to the $c_i$. The details is presented in Equation \ref{eq_hcprime} where $W$ is a matrix of weights, and $\boldsymbol{b}$ is a bias term. Each element $w_{pq}$ in $W$ is associated to the connection between the $p^{th}$ element of $v(m_j)$ and the $q^{th}$ hidden unit of the autoencoder. 

\begin{equation}
    h(v(m_j))=o(Wv(m_j)+\boldsymbol{b})
    \label{eq_hcprime}
\end{equation} 

The reconstruction $\boldsymbol{\tilde{c_i}}$ of the original word $\boldsymbol{c_i}$ by the output layer of the autoencoder is given by Equation \ref{eq_ctilde} where $W^{\prime}$ is a matrix of weights. Each element $w^{\prime}_{pq}$ in $W^{\prime}$ is associated to the connection between the $p^{th}$ hidden unit of the autoencoder and the $q^{th}$ element of the reconstruction $\boldsymbol{\tilde{c_i}}$. $\boldsymbol{b^{\prime}}$ is a bias term.
\begin{equation}
    \boldsymbol{\tilde{c_i}}=o(W^{\prime}h(v(m_j))+\boldsymbol{b^{\prime}})
    \label{eq_ctilde}
\end{equation}

\begin{figure}[ht]
    \caption{Overall architecture of the denoising autoencoder}
    \centering
    \includegraphics[scale=0.5]{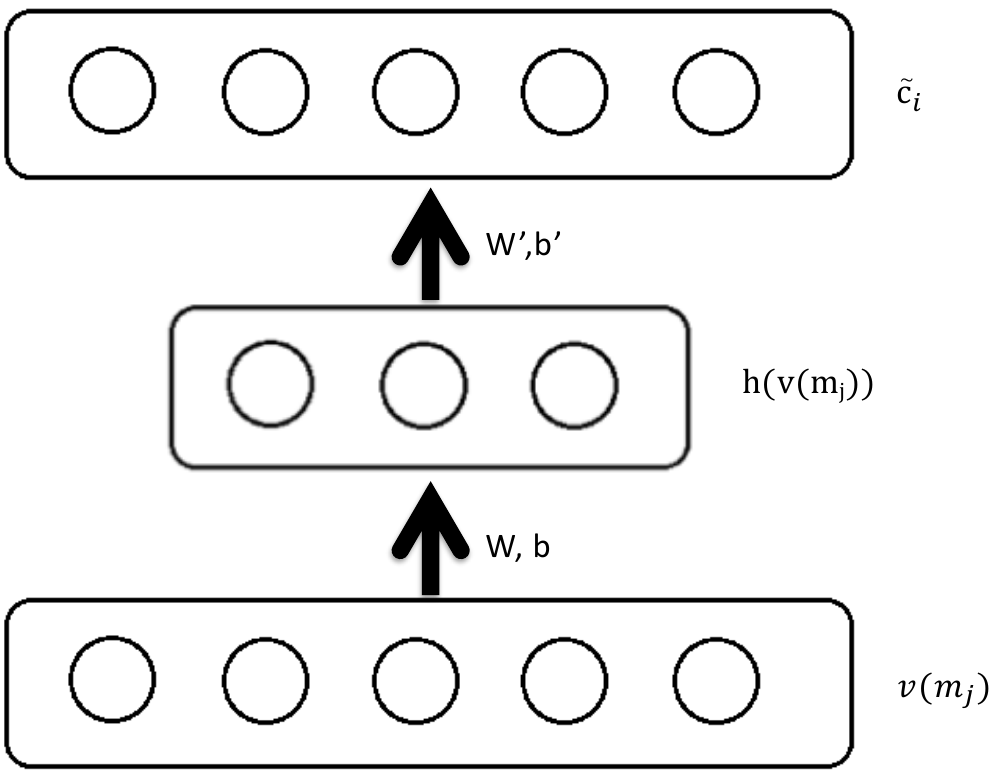}
    \label{fig:autoencoder}
\end{figure}

The overall architecture of the denoising autoencoder is presented in Figure \ref{fig:autoencoder}.

For a distance function $d$ in real vector space, the autoencoder learns the parameters $W$, $\boldsymbol{b}$, $W^{\prime}$, and $\boldsymbol{b^{\prime}}$ that minimize the loss function $L$ given by the distance between the initialization $v(m_j)$ of the non-standard version of $c_i$ and its reconstruction $\boldsymbol{\tilde{c_i}}$ (Equation \ref{eq_l}).
\begin{equation}
    L= d(\tilde{c_i},v(m_j))
    \label{eq_l}
\end{equation}

The output of the hidden layer of an encoder can be fed as an input to another autoencoder, which tries to reconstruct it. In this case, the second autoencoder learns features about the features learned by the first autoencoder: It learns a second-degree feature abstraction of the input data. This process of stacking autoencoders can be repeated indefinitely. The obtained network is called a deep believe network \cite{bengio2009learning}. For each layer, all elements in $W$, $\boldsymbol{b}$, $W^{\prime}$, or $\boldsymbol{b^{\prime}}$ are updated using back-propagation and stochastic gradient the descent \cite{lecun2012efficient}.


\subsection{Context based coding}
\label{subsec_co}

To increase the relevance of a denoising autoencoder, we connect each with their context. The context means text close to the word used in a setting.  We define the context as a sequence $a_1, a_2,..., a_T$ of words $a_t \in A$. The objective of the context based encoding is to learn a model $g$ representing the probability of a word given its context such that $g(a_t,a_{t-1},...,a_{t-s-1})=P(a_t|a_{t-1},...,a_{t-s-1})$. $g$ presents the likelihood of the word $a_t$ appearing after the sequence $a_{t-1},...,a_{t-s-1}$. This method was first introduced by Bengio et al. \cite{bengio2003neural}. We decompose the function $g$ in two parts:
\begin{enumerate}
	\item A mapping $u$ from an element $a_i \in A$ to a vector $u(a_i)$, which represents the vector associated with each word in the vocabulary. 
	\item A probability function $f$ over vector representation of words assigned by $u$ . $F$ maps an input sequence of vectors representation of words in a context, $(u(a_t),..., u(a_{t-s-1}))$, to a conditional probability distribution over words in $A$ for the next word $a_t$. Thus, $g(a_t,a_{t-1},...,a_{t-s-1}) = f(u(a_t),... ,u(a_{t-s-1}))$
\end{enumerate}

Hence, the function $g$ is a composition of the two functions, $u$ and $f$. With each of these two parts are associated some parameters. The parameters of $u$ are the elements of the matrix $U$ presenting the words vector representations. The function $f$ may be implemented by a neural network with parameters $\omega$. Training is achieved by looking for $\theta=(\omega, U)$ that maximizes the training corpus log-likelihood:
\begin{equation}
    L= \frac{1}{T} \sum_t log (g(a_t, a_{t-1}, ..., a_{t-n}))
\end{equation}

\begin{figure}[ht]
    \caption{Overall architecture of the context encoder}
    \centering
    \includegraphics[width=\textwidth]{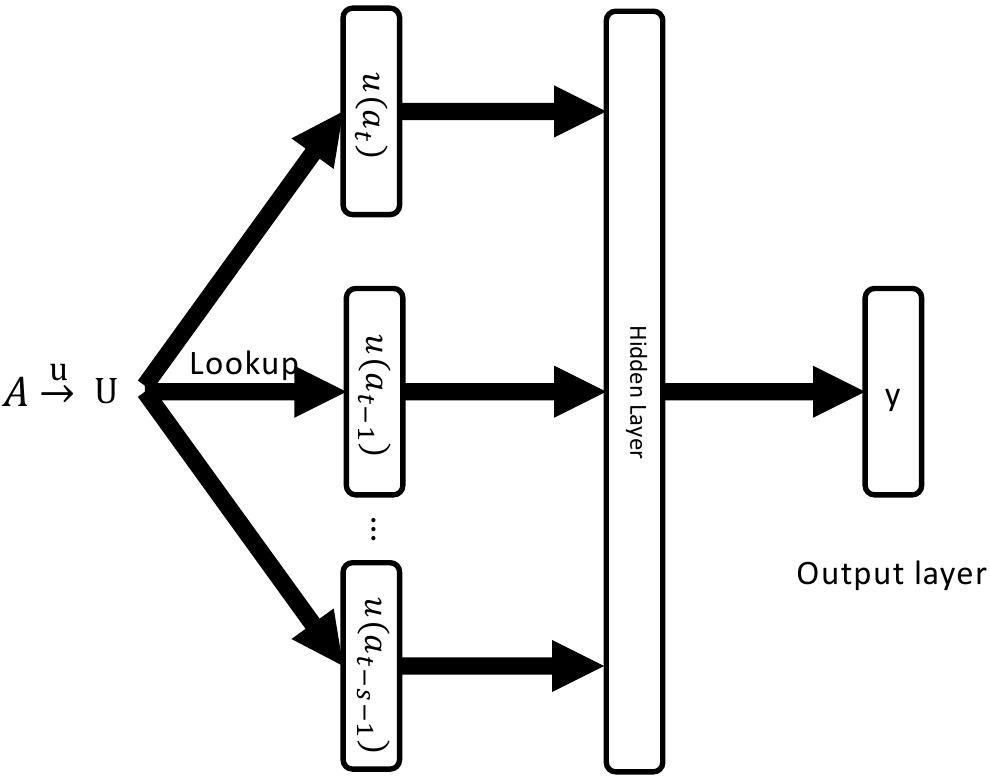}
    \label{fig:context}
\end{figure}

The neural network presenting $f$ has a softmax output layer, which guarantees positive probabilities summing to 1: 
\begin{equation}
    P(a_t|a_1, a_2,..., a_{t-1})= \frac{e^{y_i}}{\sum_j e^{y_j}}
\end{equation}
$y$ is the output of hidden layer of the neural network.

\begin{equation}
    y=\boldsymbol{\tilde{b^{\prime}}}+\tilde{W^{\prime}}o(\boldsymbol{\tilde{b}}+\tilde{W}x)
\end{equation}
where $o$ is the activation function of the hidden layer, $\tilde{W^{\prime}}$, $\tilde{W}$, and $U$ are the matrix of weights, $\boldsymbol{\tilde{b^{\prime}}}$ and $\boldsymbol{\tilde{b}}$ are the biases, and $x$ is feature vector of word vector representation from the matrix $H$: $x = (u(a_{t-1}),u(a_{t-2}),..., u(a_{t-n}))$.
The parameters of the model are $\theta={\boldsymbol{\tilde{b^{\prime}}},\boldsymbol{\tilde{b}},\tilde{W^{\prime}},\tilde{W},U}$. The overall architecture of the context encoder is presented in Figure \ref{fig:context}. 

\subsection{Distance over word space}
\label{subsec_distance}

If $f$ and $v$ are bijections, the denoising autoencoder transformation $h(v({a_i}))$ of the initialization $v(a_i)$ of a word $a_i \in C$ is a bijections from the word space $A$ to $\mathbb{R}^n$. This observation provides the function $D_a$, giving the distance between the autoencoding representations for the words $a_i$ and $a_j$ (Equation \ref{eq_D}), with the property of a metric in $A$. First, the non-negativity, triangular inequality, and symmetry of $D$ are derived from the same properties of the $d$. Second, the identity of indiscernible is automatically deduced from the observation that $h$ and $v$ are bijective functions. The advantage of this distance is that the function $h$ contains in its weight matrix $W$ and bias $\boldsymbol{b}$ an encoding that captures the stochastic structure of misspelling patterns representing the words observed during the learning phase of the autoencoder. It is important to note that $D_a$ is a pseudometric on $A$ since the purpose of the autoencoder is to minimize the distance between vector representation of a correct word and its non-standard version so the identity of indiscernibles cannot be guaranteed.  

\begin{equation}
    D_a(a_i, a_j)= d(h(v(a_i)),h(v(a_j)))
    \label{eq_D}
\end{equation}

\begin{figure}[ht]
    \caption{Overall architecture of the autoencoder in combination with the context encoder to find the similarity between the words }
    \centering
    \includegraphics[width=\textwidth]{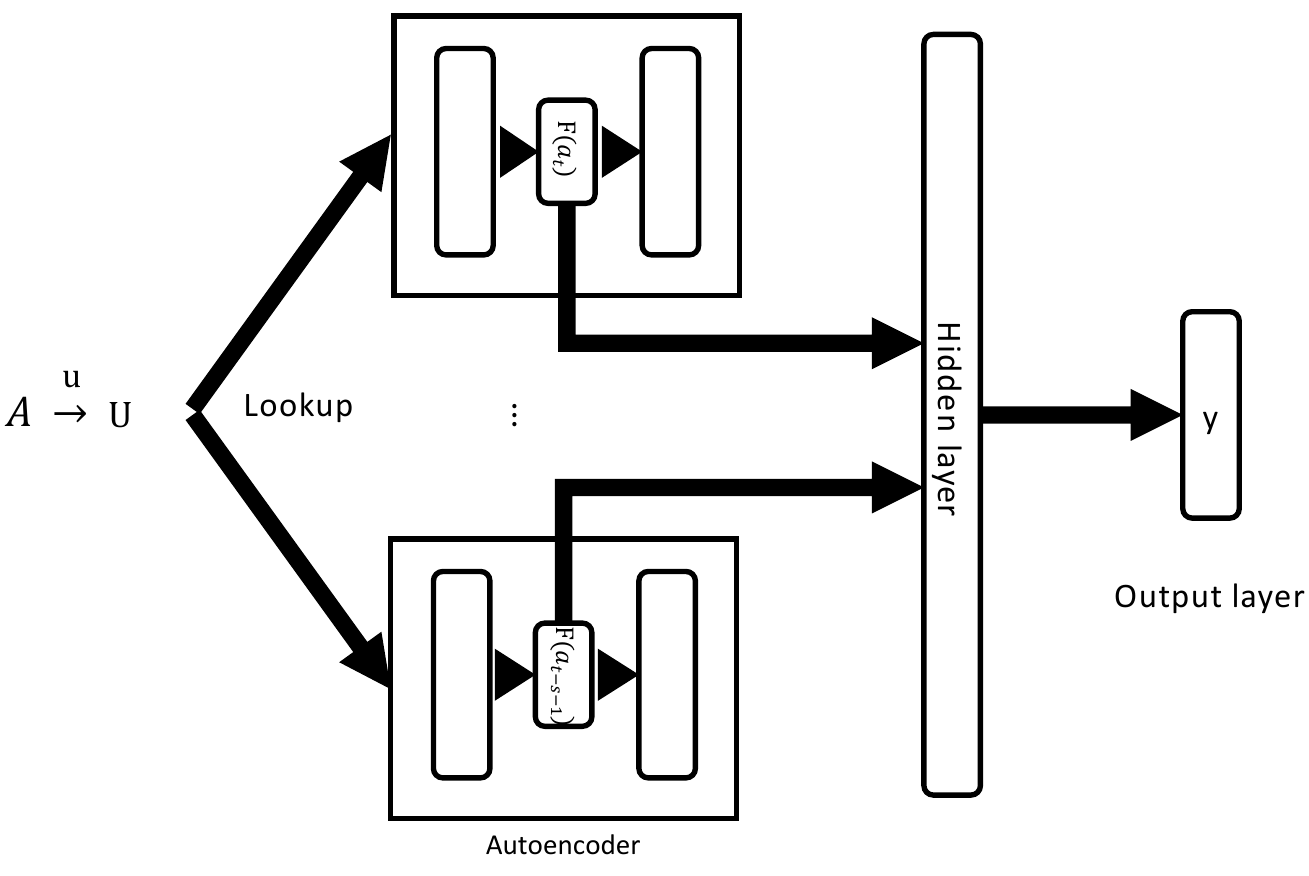}
    \label{fig:achitec}
\end{figure}

To get the mapping $F$, the Matrix $U$ in section \ref{subsec_co} can be updated by combining the autoencoder in section \ref{subsec_do} and the context encoder. In such case, both methods will work in a parallel manner to update the vector representation of the words.
Thus, the vector representations of the words in the matrix $U$ are learned using the context method, and the vector representations of non-standard words in $U$ are calculated based the autoencoder in addition to that (see Figure \ref{fig:achitec}). The denoising autoencoder we used in, in this case, is a seven-layer deep autoencoder. The initialization function $v$ is a one-hot encoding of the words in $A$ which results in an input layer of the denoising autoencoder of $|A|$ nodes (not shown in Figure \ref{fig:achitec} for the sake of presentability). The combination of the autoencoder and the context coding to produce the mapping $F$ is used to define the function $D_c$ in $A$ in Equation \ref{eq_D1}. $D_c$ can be seen as an extension of $D_a$, but instead including the context of the words. $D_c$ is a function that finds the distance between the words $a_i$ and $a_j$ using the function $F$. $D_c$ is not a metric in $A$ for the same reason $D_a$ is not, and it is not a metric in $C$ as well since the context encoding will give words used in the same context similar vector presentation, therefore, the identity of indiscernibles is not guaranteed here either. 
\begin{equation}
    D_c(a_i, a_j)= d(F(a_i),F(a_j))
    \label{eq_D1}
\end{equation}

\section{Results and discussion}
\label{sec_results}

To test our approach, we used a data set composed of the 1051 most frequent words from Twitter paired with their various misspellings, from a data set was initially used in an IBM data normalization challenge in 2015 \footnote{The data is available here: \url{https://noisy-text.github.io/norm-shared-task.html}}. To train the context based encoder, we used a data set containing 97191 different sentences with vocabulary words and their non-standard form. It is important to note that the data is imbalanced: Some words have only one non-standard form, and other words have multiple non-standard forms. This imbalance may potentially introduce some challenges since the autoencoder might not learn accurate features for words with a few non-standard versions. Since we used softmax units, the number of nodes in the last hidden layer of the autoencoder is set to $11  \simeq log_2(1051)$, which also represents the length of the obtained encoding vector. The autoencoder is trained using min-batch gradient descent with batches of 100 examples each and a learning rate of 0.01. The closest standard word is picked as the most likely standard version of the non-standard spelling. By closest, we mean the word that has the smallest distance $D_a$ or $D_c$ to the non-standard spelling as defined in subsection\footnote{The code and data part of these experiments are available here \url{https://github.com/mehdi-mbl/WordCoding}} \ref{subsec_distance}.

\begin{table*}[!t]
    \caption{Performance comparison including our two approaches: $D_a$ Denoising autoencoder and $D_c$ the ombination of autoencoder and context encoder}
    \label{table_performance}
        \centering
        \resizebox{\textwidth}{!}{
        \begin{tabular}{|c|c|c|}
        \hline
        Distance & Closest word & 5th closest word\\
        \hline
        Cosine Similarity & 46.33\% & 60.22\%\\ 
       \hline
        Q-Gram & 47.57\% & 62.41\%\\ 
        \hline
        Sørensen-Dice coefficient & 47.85\% & 60.03\%\\

        \hline
        Edit distance & 55.75\% & 68.22\%\\
        
        \hline
        Weighted-Levenshtein & 55.85\% & 67.93\%\\

        \hline
        Damerau-Levenshtein distance & 56.51\% & 68.03\%\\ 
        \hline
        N-Gram & 58.23\% &  76.49\%\\ 
        \hline
        Metric-Longest Common Subsequence & 60.89\% & 75.73\%\\ 
        \hline
        Longest Common Subsequence & 61.37\% & 74.31\%\\

        \hline
        Normalised-Levenshtein & 63.17\% & 78.30\%\\

        \hline
        $D_a$ with Cosine similarity & 83.82\% & 89.53\%\\
       \hline
        $D_c$ with $L_1$ distance & 76.37\% & 81.53\%\\
 
        \hline
        $D_c$ with Euclidean distance & 82.71\% & 87.35\%\\ 
        \hline
       $D_c$ with Cosine similarity & \textbf{85.37\%} & \textbf{89.61\%}\\
        \hline
    \end{tabular}}
\end{table*}

\begin{figure}[ht]
    \caption{Performance of different distances in finding the standard form of a non-standard word}
    \centering
    \includegraphics[scale=0.5]{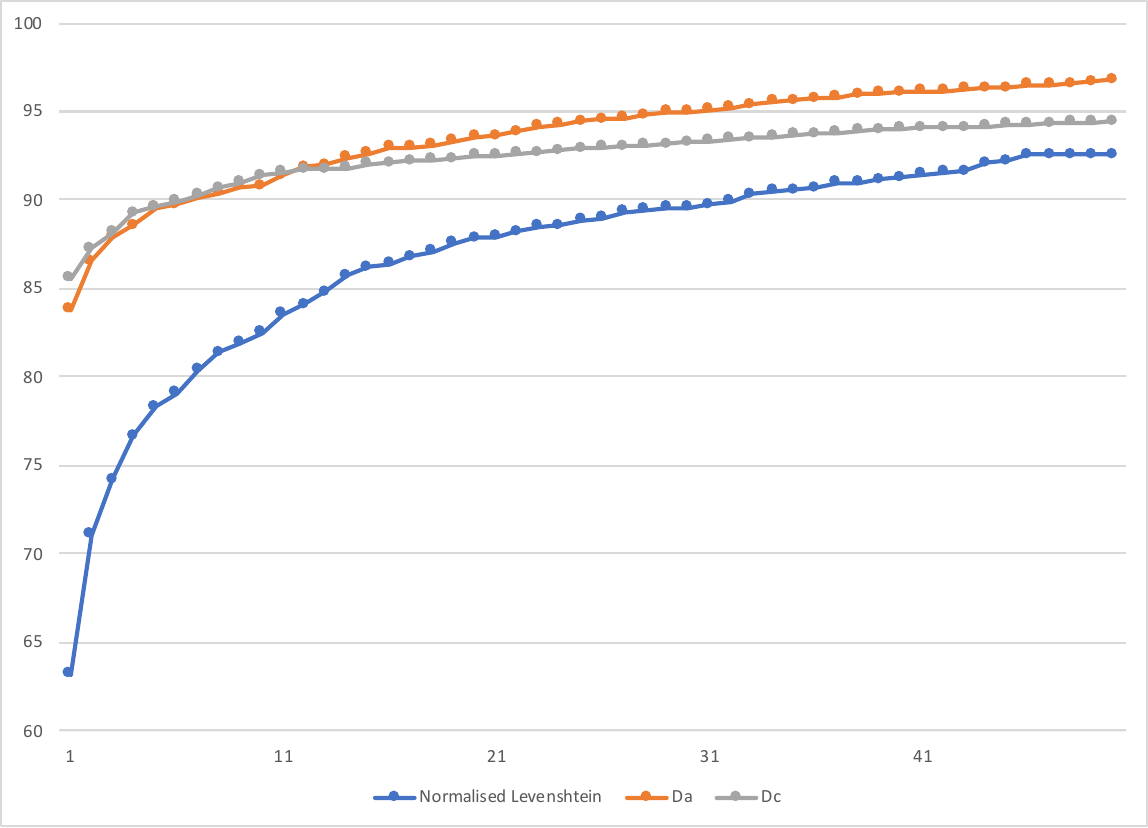}
    \label{fig:performance}
\end{figure}

Table \ref{table_performance} compares the results produced our approaches $D_a$, $D_c$, with the existing string metrics presented in Section \ref{sec_background} in finding the correct version of a non-standard spelling. The table shows a huge increase in accuracy from 63.17\% for the best metric available in the literature (Normalised-Levenshtein) to 85.37\% when $D_c$ is used. The reason is that unlike the state-of-the-art metric, $D_c$ captures stochastic word patterns shared between the nonstandard word and its correct form. Figure \ref{fig:performance} shows the performance of the Normalised-Levenshtein, $D_a$, and $D_c$ in finding the standard spelling of a non-standard word among its nearest neighbors. The $x$ axis represents the number of nearest neighbors. Figure \ref{fig:performance} shows that after ten neighbors $D_a$ starts to outperform $D_c$ because $D_a$ is modeled by an autoencoder which the main purpose is to model such non-standard word. With $D_c$, as we go farther from a word, the nearest word will contain words used in a similar context which are not necessarily standard version of the word (see table \ref{table_nwsfc}).

Our approach is not limited to one vector distance. In fact, neural representation part of the inner autoencoder can be measured with any vector similarity. 

Table \ref{table_performance} also compares the performance of $D_c$ with different vector distances. The cosine similarity produces the best performance in this task with 85.37\%.

Table \ref{table_nwsf} shows the closest correct word to a sample of non-standard spellings with the autoencoder $D_a$ without context encoder. Table \ref{table_nwsf} shows the results that the closest words share some patterns with the non-standard counterpart. For example, the four closest word to ``starin" all end with ``ing" and three of them start with ``s" (staring, praying sucking, slipping). Notice also the similarity in character between the two closest words ``staring" and ``praying". The same thing can be said about the closest word to ``ddnt". In the case of ``omg" and ``justunfollow", all the closest words are combinations of more than one word, which suggests that the autoencoder learns that they are abbreviations or combination of words. The next examples in Table \ref{table_nwsf} presents a non-standard spelling for which the approach with denoising autoencoder fails to recognize the correct version in the five closest word: The correct version of ``wada" is ``water" but our algorithm chooses ``wanna" as the closest correct version. Even though it is not the correct guess, the resemblance between ``wada" and ``wanna" justifies the guess and a human could arguably have made the same mistake. The same can also be said about ``bea" and ``the". For ``dats", the algorithm picks the correct word as the fourth closest word. However, the first pick (``what's") can also be justified since it shares three characters with the non-standard spelling.

\begin{table*}[!t]
    \caption{Example of words and their closest standard form using denoising autoencoder}
    \label{table_nwsf}
    \centering
    \resizebox{\textwidth}{!}{
    \begin{tabular}{|c|c|c|c|c|c|c|}
        \hline
        Non-standard spellings & Closest word & 2nd closest word & 3rd closest word & 4th closest word & 5th closest word & Correct word\\
        \hline
        thng & \textbf{thing} & there & wanna & right & where & thing\\
        \hline
        starin & \textbf{staring} & praying & sucking & slipping & weekend & staring\\
        \hline
        omg	& \textbf{oh my god} & at least & in front & in spite & what's up & oh my god\\
        \hline
        ddnt & \textbf{didn't} & that's & aren't & what's & better & didn't\\
        \hline
        justunfollow & \textbf{just unfollow} & what about you & ultra violence & direct message & what are you doing & just unfollow\\
        \hline
        wada & wanna & sugar & sense & never & speed & water\\
        \hline
        dats & what's & wasn't & aren't & \textbf{that's} & give a & that's\\
        \hline
        bea & the & why & kid & old & yes & tea\\
        \hline
    \end{tabular}}
\end{table*}

Table \ref{table_nwsfc} shows the closest words to a sample of other words in term of the distance $D_c$. In addition to the non-standard spelling being closed to the standard word, Table \ref{table_nwsfc} shows that words similar in meaning are also introduced to the closest words. For example, the third closest word to ``dogg" is  ``cat" both being domestic animals. Notice also that ``boy" and ``kid" come next because in many of the training sentence a boy a kid is mentioned in the same context as a dog or cat. The same thing can be said about the closest word to ``tomorrow". In the case of ``txt" and ``birthday", most of the closest words are their standard/ non-standard version. The next examples in Table \ref{table_nwsfc} presents a non-standard spelling, ``teering", for which the approach with the distance $D_c$ fails to recognize a word with a close meaning to it in the five closest word. The correct version of ``teering" is ``tearing" which is the third closest word in table \ref{table_nwsfc}. Even though the closest word to ``teering" is not related to it, the resemblance between ``weering" and ``teering" can justify it being the closest word. Table \ref{table_nwsfc} shows that $D_c$ finds the standard version of ``bea" as the thirds nearest word which is an improvement in this case over $D_a$.

\begin{table*}[!t]
    \caption{Example of words and their closest standard form using combination of denoising autoencoder and context coding}
    \label{table_nwsfc}
    \centering
    \resizebox{\textwidth}{!}{
    \begin{tabular}{|c|c|c|c|c|c|c|c|}
        \hline
        Non-standard spellings & Closest word & 2nd closest word & 3rd closest word & 4th closest word & 5th closest word \\
        \hline
        dogg & dog & doog & cat & boy & kid \\
        \hline
        txt & text & texting &  txted  & texted & work \\
        \hline
        teering	& wearing & meeting & tearing & shaking & picking \\
        \hline
        tomorrow & tmrw & tmr & today & yesterday & judgment\\
        \hline
        video & vid & vids & videos & sleep & remix\\
        \hline
        birthday & bday & biryhday & birthdayyy & drinking & dropping\\
        \hline
        thng & ting & thing & think & right & stuff \\
        \hline
        starin & staring & looking & glaring & slipping & praying\\
        \hline
        omg	& omgg & omfg & ohmygod & ohmygad & oh my god\\
        \hline
        ddnt & didn & didnt & didn't &  havn't & aren't\\
        \hline
        wada & wanna & sugar & sense & never & speed \\
        \hline
        dats & dat & that & thats & thts & that's \\
        \hline
        bea & the & tea & yes & old & coffee\\
        \hline
    \end{tabular}}
\end{table*}

\begin{figure}[ht]
    \caption{2D plot of some example of vector representation of words using T-SNE}
    \centering
    \includegraphics[width=\textwidth]{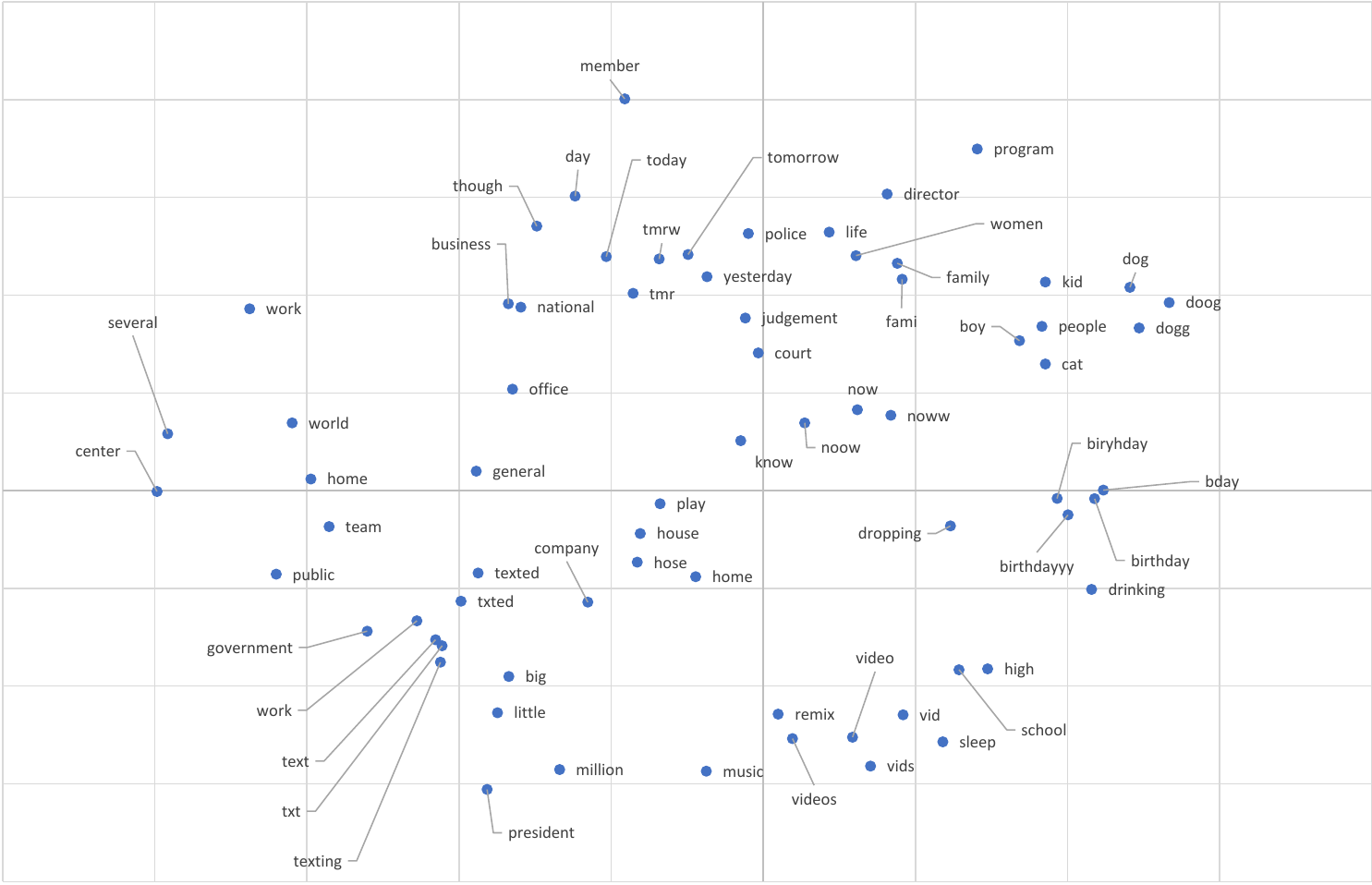}
    \label{fig:2d}
\end{figure}

Figure \ref{fig:2d} shows a scatter plot of the vector representation of the words presented in table \ref{table_nwsfc} after a dimension reduction. T-SNE \cite{maaten2008visualizing} was use for the dimension reduction.

\section{Conclusion}
\label{sec_conclusion}
In this paper, we proposed to combine a denoising autoencoder along with a context encoder to learn a mapping between vocabulary and real vector space. This mapping allows us to define a metric in word space that encompasses non-standard spelling variations of words as well as words used in similar context. This work is a first attempt at defining a fully learned metric in word space using neural networks. Granted, the resulting metric does not satisfy all the theoretical property of a metric. However, the experimental results showed that the resulting metrics succeeds in 85.4\% of the cases in finding the correct version of a non-standard spelling --- a considerable increase in accuracy from the established Normalised Levenshtein of 63.2\%. Besides, we showed that words used in similar contexts have a shorter distance between them than words in different contexts. 

\appendix
\section{Model's parameters and functions}
\label{apx_pandf}

\begin{table}[H]
    
    \label{table_parameters}
        \centering
        \begin{tabular}{|c|p{8cm}|}
        \hline
        Parameters/functions & Description\\
        \hline
        $A$ & Vocabulary\\ 
        \hline
        $C$ & Set of standard words in $A$\\
        \hline
        $M_{c_i}$ & Set of non-standard version of the word $c_i \in C$\\
        \hline
        $v$ & Initialization function $A \to \mathbb{R}^n$\\
        \hline
        $h$ & Output of the hidden layer of the denoising autoencoder\\
        \hline
        $W$, $W^{\prime}$& Weights of the denoising autoencoder\\
        \hline
        $\boldsymbol{b}$, $\boldsymbol{b^{\prime}}$ & Biases of the denoising autoencoder\\
        \hline
        $\tilde{c_i}$ & Output of the denoising autoencoder\\
        \hline
        $L$ & Loss function\\
        \hline
        $d$ & Metric in real vector space\\
        \hline
        $o$ & Activation function\\
        \hline
        $g$ & Probability distribution over $A$\\
        \hline
        $u$ & Mapping function $A \to \mathbb{R}^n$\\
        \hline
        $f$ & Probability distribution over mappings of words produced by $u$\\
        \hline
        $y$ & Output of the context encoder\\
        \hline
        $\tilde{W}$, $\tilde{W^{\prime}}$ & Weights of the context encoder\\
        \hline
        $\boldsymbol{\tilde{b}}$, $\boldsymbol{\tilde{b^{\prime}}}$ & Biases of the context encoder\\
        \hline
        $F$ & Mapping function $A \to \mathbb{R}^n$ learned by the combination of the denoising autoencoder and context encoder\\
        \hline
        $D_a$ & Metric over $A$ resulting from the denosing entoencoder mapping \\
        \hline
        $D_c$ & Metric over $A$ resulting from $F$\\
        \hline
    \end{tabular}
\end{table}

\section*{References}
\bibliographystyle{elsarticle-num}
\bibliography{Bibliography.bib}
\end{document}